\documentclass[prd,nofootinbib,preprint,superscriptaddress]{revtex4}

\usepackage{graphicx}

\usepackage[hypertex]{hyperref}
\newcommand{\beq}{\begin{equation}}
\newcommand{\eeq}{\end{equation}}

\newcommand{\capdef}{}
\newcommand{\mycaption}[2][\capdef]{\renewcommand{\capdef}{#2}%
       \caption[#1]{{\footnotesize #2}}}
\makeatletter
\renewcommand{\fnum@table}{\textbf{\tablename~\thetable}}
\renewcommand{\fnum@figure}{\textbf{\figurename~\thefigure}}
\makeatother

\begin{document}
\pagestyle{plain}

\vspace*{1cm}
\preprint{CERN-PH-TH/2008-103}
\preprint{VPI-IPNAS-08-10}

\title{A low energy neutrino factory with non-magnetic detectors
\vspace*{1cm}}

\author{\bf Patrick Huber}
\email{pahuber_at_vt.edu}
\affiliation{Physics Department, Theory Division, CERN,
1211 Geneva 23, Switzerland}
\affiliation{Department of Physics, Virginia Tech, Blacksburg, VA
  24062, USA \vspace*{1cm}}

\author{\bf Thomas Schwetz}
\email{schwetz_at_cern.ch}
\affiliation{Physics Department, Theory Division, CERN,
1211 Geneva 23, Switzerland}


\begin{abstract}

  \vspace*{5mm} We show that a very precise neutrino/anti-neutrino
  event separation is not mandatory to cover the physics program of a
  low energy neutrino factory and thus non-magnetized detectors like
  water Cerenkov or liquid Argon detectors can be used. We point out,
  that oscillation itself strongly enhances the signal to noise ratio
  of a wrong sign muon search, provided there is sufficiently accurate
  neutrino energy reconstruction. Further, we argue that apart from a
  magnetic field, other means to distinguish neutrino from
  anti-neutrino events (at least statistically) can be explored.
  Combined with the fact that non-magnetic detectors potentially can
  be made very big, we show that modest neutrino/anti-neutrino
  separations at the level of 50\% to 90\% are sufficient to obtain
  good sensitivity to CP violation and the neutrino mass hierarchy for
  $\sin^22\theta_{13}>10^{-3}$. These non-magnetized detectors have a
  rich physics program outside the context of a neutrino factory,
  including topics like supernova neutrinos and proton decay. Hence,
  our observation opens the possibility to use a multi-purpose
  detector also in a neutrino factory beam.

\end{abstract}
\maketitle


\section{Introduction}

A neutrino factory is a neutrino source based on the decay of muons
stored in a decay ring with long straight sections~\cite{Geer:1997iz}.
The muons are moving at relativistic speed in the decay ring and
hence, the isotropic decay in their rest frame becomes a highly
collimated beam in the laboratory system. The neutrino beam consists
for the decay of $\mu^+$, assuming no net muon polarization, of equal
numbers of $\bar\nu_\mu$ and $\nu_e$. The resulting charged current (CC)
muon signals in the detector are, schematically,
\begin{equation}
\mu^+ {\nearrow\atop\searrow} \begin{array}{c}
\bar\nu_\mu\stackrel{P_{\bar\mu\bar\mu}}{\longrightarrow}
\bar\nu_\mu\stackrel{\sigma^{\bar\mu}_\mathrm{CC}}{\longrightarrow}\mu^+
\\[6mm]
\nu_e  \stackrel{P_{e\mu}}{\longrightarrow}
\nu_\mu\stackrel{\sigma^{\mu}_\mathrm{CC}}{\longrightarrow}\mu^-  
\end{array}\,.
\end{equation}
The appearance signal due to the oscillation probability $P_{e\mu}$ is
thus proportional to the number of $\mu^-$ events, which have the
opposite sign with respect to the initial decaying $\mu^+$ and
therefore are called ``wrong sign'' muon events, in contrast to the
``right sign'' muons from disappearance channel, appearing for a
non-vanishing survival probability $P_{\bar\mu\bar\mu}$ for the
$\bar\nu_\mu$. Of course, the analogous relations hold for $\mu^-$
decaying. Throughout this letter, whenever we talk about $\mu^+$ in
the storage ring, the CP analogous channel stemming from $\mu^-$
stored is implied, unless otherwise mentioned.  In a traditional
neutrino factory with energies around $25\,\mathrm{GeV}$ of the
decaying muons one uses a magnetized iron calorimeter and the
resulting curvature of the muon track to identify the muon charge with
backgrounds at the $10^{-4}-10^{-3}$ level, which is the key to the
extraordinary sensitivity of a neutrino factory to even small values
of $P_{e\mu}$.  For a current, comprehensive review,
see~\cite{issacc:2008xx,Abe:2007bi,Bandyopadhyay:2007kx}.

It has been realized, however, that a traditional neutrino factory
does not perform very well for large values of
$\sin^22\theta_{13}>10^{-2}$ and therefore, a so-called ``low energy''
neutrino factory has been proposed~\cite{Geer:2007kn,Bross:2007ts}
with a muon energy of around $5\,\mathrm{GeV}$, see
also~\cite{Huber:2007uj}. At those energies, muon tracks in iron are
too short to allow a unique determination of the curvature and thus
charge. The solution put forward in~\cite{Geer:2007kn,Bross:2007ts} is
to use a totally active scintillator detector (TASD), like
MINERVA~\cite{Minerva} immersed in a magnetic field of about
$0.5\,\mathrm{T}$. Preliminary simulations presented
in~\cite{Geer:2007kn,Bross:2007ts} indicate that the performance of
such a magnetized TASD is satisfactory. However the very large number
of readout channels and the need to magnetize a large volume make it
difficult to scale this detector to fiducial masses much larger than
$10-20\,\mathrm{kt}$.

In this work we will demonstrate that a very precise charge
identification is not mandatory to cover the physics program of a low
energy neutrino factory and thus non-magnetized detectors like water
Cerenkov (WC) or liquid Argon (LAr) detectors can be used (see
also~\cite{Freund:2000ti, Aoki:2003kc}). We argue that apart from a
magnetic field, other means to distinguish neutrino from anti-neutrino
events (at least statistically) can be explored. Combined with the
fact that such detectors potentially can be made very big, we show
that modest charge identification abilities (at the level of 50\% to
90\%) are enough to be competitive with the above mentioned magnetized
TASD detector. These non-magnetized detectors have a vast physics
program outside the context of a neutrino factory, including topics
like supernova neutrinos and proton decay. For a recent review,
see~\cite{Autiero:2007zj}. Hence, our observation opens the attractive
possibility to use a multi-purpose detector also in a neutrino factory
beam.

The outline of the paper is as follows. In section~\ref{sec:principle}
we show that oscillations by themselves suppress the background of
wrong sign muons, and therefore, in principle even without any charge
identification there is some sensitivity to the appearance signal.  In
section~\ref{sec:cid} we discuss some means to separate neutrino and
anti-neutrino events without using a magnetic field and we introduce a
simple (idealized) parametrization to describe statistically
neutrino/anti-neutrino--enhanced data samples. In
section~\ref{sec:sensitivities} we present the results of sensitivity
calculations for CP violation and the neutrino mass hierarchy,
comparing non-magnetized detectors with some modest
neutrino/anti-neutrino separation abilities to the reference
magnetized TASD. We conclude in section~\ref{sec:conclusions}.

\section{A neutrino factory without charge identification}
\label{sec:principle}

The central observation, this paper is based on, is that the $\nu_\mu$
from the disappearance channel, which give rise to the so-called right
sign muons, will have almost completely turned into $\nu_\tau$ for
energies around the first oscillation maximum, which we denote by
$E_\mathrm{1st}$ and is defined by $\Delta \equiv \Delta
m^2_{31} L/(4E) = \pi/2$. For exactly maximal mixing, {\it i.e.}\
$\theta_{23}=\pi/4$, the survival probability $P_{\mu\mu}$ becomes
practically zero at $E_\mathrm{1st}$ and stays small within a narrow
energy range centered on $E_\mathrm{1st}$: 
\begin{equation}\label{eq:surv}
P_{\mu\mu} = 1 - \sin^22\theta_{23}\sin^2\Delta + 
\mathcal{O}(\Delta m^2_{21}, \theta_{13}) \,.
\end{equation}
On the other hand, the appearance probability $P_{e\mu}$ leading to
the wrong sign muons will peak around $E_\mathrm{1st}$. In vacuum, for
simplicity, one has 
\begin{equation}\label{eq:app}
P_{e\mu} \approx
4 s_{13}^2 s_{23}^2 \sin^2\Delta + 
2 \tilde\alpha s_{13} \sin2\theta_{23} \sin\Delta \cos(\Delta \mp \delta)
+ c_{23}^2 \tilde\alpha^2 \,,
\end{equation}
with $s_{ij} \equiv \sin\theta_{ij}$, $c_{ij} \equiv \cos\theta_{ij}$,
$\tilde\alpha \equiv \sin2\theta_{12} \, \Delta m^2_{12}L/(4E)$, and
'$-$' ('$+$') holds for neutrinos (anti-neutrinos). Thus using events
in the region around $E_\mathrm{1st}$ a reasonable signal to noise
ratio can be obtained even if there is no possibility to distinguish
neutrino from anti-neutrino events. Therefore, a good energy
resolution of the detector will be crucial to maximally exploit the
suppression of right sign muons due to oscillation. At the typical
energies of a low energy neutrino factory of a few GeV the
contribution of quasi-elastic scattering is still large enough to
provide sufficient energy resolution without the need of accurate
hadronic calorimetry.

\begin{figure}
\centering \includegraphics[width=0.9\textwidth]{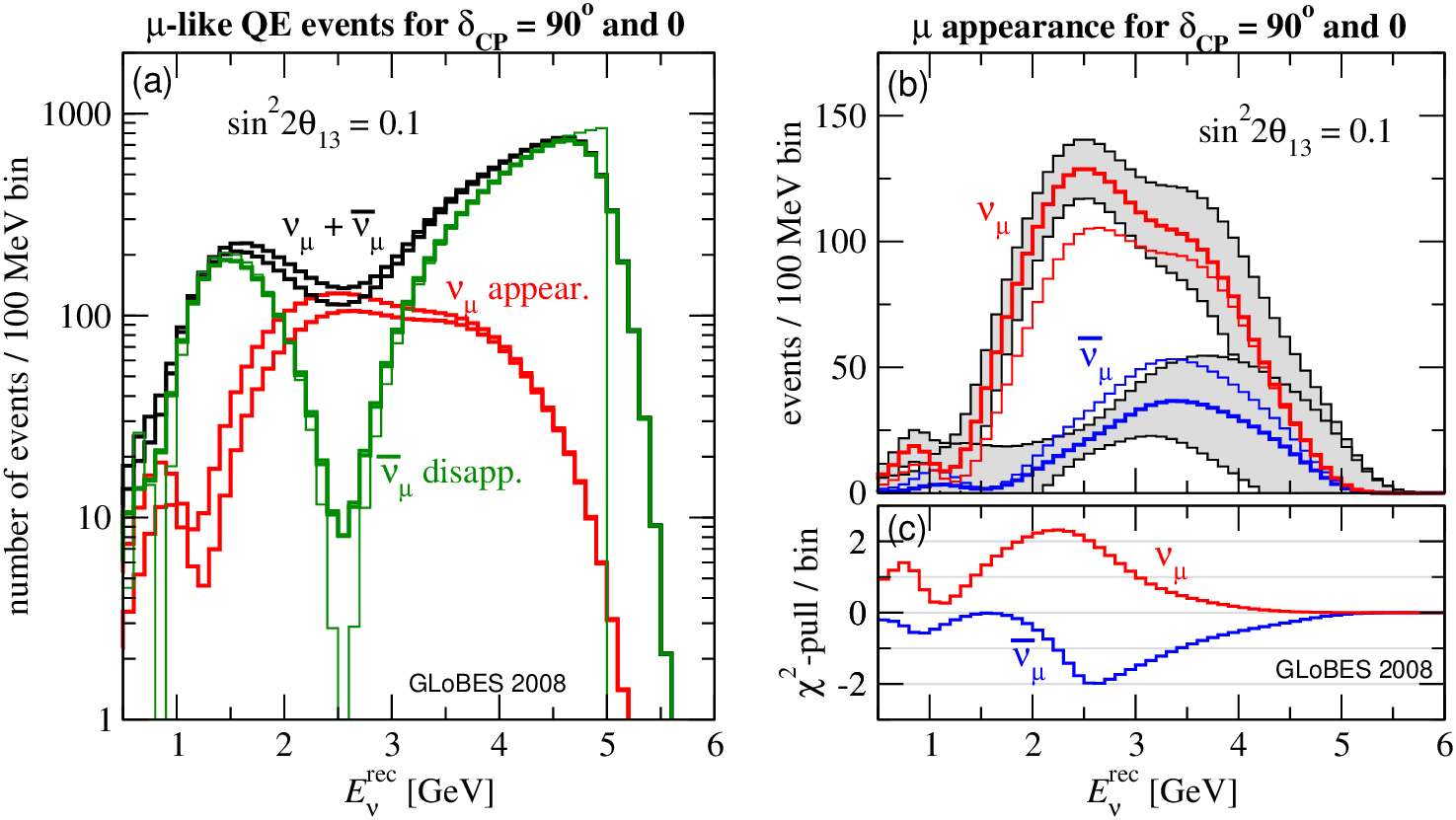}
  \mycaption{\label{fig:spec} Event rate spectra for
  $\sin^22\theta_{13}=0.1$ for quasi-elastic charged current events in
  a LAr detector as described in table~\ref{tab:det}.  For panel~(a)
  we assume stored $\mu^+$ and show the right sign muon events
  (``$\bar\nu_\mu$ disapp.''), the wrong sign muon events (``$\nu_\mu$
  appear.'') and the sum of all muon events (``$\nu_\mu +
  \bar\nu_\mu$''). The upper thick lines are for $\delta=+90^\circ$
  and the lower ones are for $\delta=0^\circ$. The thin line shows the
  right sign muon events in the case of perfect energy
  resolution. Panel~(b) shows the background subtracted wrong sign
  events for stored $\mu^+$ (``$\nu_\mu$'') and for stored $\mu^-$
  (``$\bar\nu_\mu$'') with their resulting $1\,\sigma$ error bars
  (gray shaded regions) for $\delta = +90^\circ$. Thin lines
  correspond to $\delta=0^\circ$. Panel~(c) shows the significance per
  bin in the difference between $\delta = +90^\circ$ and $0^\circ$.}
\end{figure}

Figure~\ref{fig:spec} shows event rate spectra expected in a 100~kt
liquid Argon detector at a distance of 1290~km from a neutrino
factory.  The energy of the stored muons is $5\,\mathrm{GeV}$ and we
assume a total of $10^{22}$ useful muon decays, equally divided into
$\mu^-$ and $\mu^+$ running. The events shown are quasi-elastic events
and we assume $\sin^22\theta_{13}=0.1$.
Figure~\ref{fig:spec}(a) corresponds to $\mu^+$ decays and shows the
spectra for right sign muons ($\bar\nu_\mu$ disappearance) and wrong
sign muons ($\nu_\mu$ appearance), as well as the sum of all muon
events. First we observe, that in the energy region from about
$2-3\,\mathrm{GeV}$ the wrong sign signal exceeds the right sign
background; the maximal signal to background ratio is about 10, which
happens at approximately $E_\mathrm{1st}$. This proves that
oscillation on its own provides an effective mechanism to suppress the
right sign muon background to a wrong sign muon search. The thick
lines do include an energy resolution of $\Delta E = 0.05\sqrt{E} +
0.085$ in units of GeV, whereas the thin line shows the right sign
muon background in the case of perfect energy reconstruction. The
effect of a finite energy resolution is to move events into the
oscillation dip and thereby to increase the background for the wrong
sign muon signal.

In principle, the $\nu_\tau$ resulting from
$\nu_\mu\rightarrow\nu_\tau$ oscillations can give rise to right sign
muons as well, for those cases where the $\tau$ lepton from a charged
current interaction decays leptonically into a muon. The branching
fraction for this decay mode is only about 17\%~\cite{PDBook}.
Moreover, there is strong suppression of the charged current cross
section due the finite mass of the $\tau$
lepton~\cite{Paschos:2001np}. We have estimated that a total of
$\sim600$ $\nu_\tau$ charged current events would be obtained in
$100\,\mathrm{kt}$ detector mass. Of these, only 17\% would produce a
right sign muon, {\it i.e.} about 100 events.  Assuming that the tau
lepton carries all the energy of the parent $\nu_\tau$, we can compute
the resulting muon spectrum. The result is about 10 events per bin in
the peak of their distribution, which however happens at energies well
below $E_\mathrm{1st}$. Thus right-sign muons from tau decay never
make up more than a few percent of the right-sign muons from genuine
$\nu_\mu$ charged current events in the relevant energy range.
Therefore, these events are not included in our analysis. Note, that
these numbers depend sensitively on the chosen muon energy in the
storage ring, since the $\nu_\tau$ events stem exclusively from the
high energy part of the neutrino spectrum from $4-5\,\mathrm{GeV}$;
thus a decrease in muon energy to $4\,\mathrm{GeV}$ would virtually
eliminate the $\nu_\tau$ events, whereas an increase to
$6\,\mathrm{GeV}$ would lead to 6-fold increase in $\nu_\tau$ events.

Figure~\ref{fig:spec}(a) displays two sets of thick lines: the upper
set of lines is computed for $\delta=90^\circ$, whereas the lower set
of curves is computed for $\delta=0^\circ$. We observe, that the right
sign muon signal exhibits only a very weak dependence on the value of
$\delta$, which is crucial in order to allow for a clean extraction of
CP effects. As a result, the full dependence on $\delta$ shown by the
wrong sign muons is preserved in the sum of both signs of muons.
Figure~\ref{fig:spec}(b) shows the background subtracted appearance
signal event spectra. The gray bands depict the resulting statistical
$1\,\sigma$ errors, which are computed from the sum of right and wrong
sign events. This is shown for $\mu^+$ stored ($\nu_\mu$ appearance)
and for $\mu^-$ stored ($\bar\nu_\mu$ appearance). The thick lines are
for $\delta=90^\circ$, whereas the thin lines are for
$\delta=0^\circ$. We see, that in the bins with the best signal to
noise ratio, each bin provides around $2\,\sigma$ of significance as
shown in panel~(c).  We also see that the effect goes in opposite
directions for neutrinos and anti-neutrinos thus manifestly displaying
CP violation. This remains true if also the second CP conserving case,
$\delta = 180^\circ$, is taken into account. Note that one can even
discern the effects from the second oscillation maximum around
$1\,\mathrm{GeV}$. 

The discussion so far has assumed maximal mixing $\theta_{23}=\pi/4$.
From equation~\ref{eq:surv} follows that if $\theta_{23} \neq \pi/4$
the survival probability $P_{\mu\mu}$ will not go to zero at the
first oscillation maximum and therefore somewhat more wrong sign muons
will end up in the signal region around $E_\mathrm{1st}$.
Nevertheless, as we will show in section~\ref{sec:sensitivities}, for
values of $\theta_{23}$ within the currently allowed $2\,\sigma$ range
the suppression of wrong sign events around $E_\mathrm{1st}$ is still
sufficient and does not alter our results significantly.

\section{Neutrino/anti-neutrino separation without a magnetic field}
\label{sec:cid}

Neutrino and anti-neutrino quasi-elastic (QE) charged current events
differ by a number of obvious and also more subtle signatures. The
reactions are given by
\begin{equation}\label{eq:QE}
\nu_x + N \rightarrow l_x^-+p+N'
\quad\mathrm{and}\quad
\bar\nu_x + N \rightarrow l_x^++n+N'\,,
\end{equation}
where $l_x$ denotes a charged lepton with $x$ being $\mu$ or $e$ and
$N$ is the nucleus. A traditional neutrino factory experiment aims at
measuring the charge sign of the outgoing lepton $l_x$ by using a
magnetic field and the resulting curvature of the track. This
technique, currently, is planned to be applied only to muons, since
electron tracks are considered neither long nor clean enough. In the
following we mention three other signatures which can be used in
principle to distinguish neutrino from anti-neutrino events without
using a magnetic field, where we do not exclude that in a specific
detector additional signatures beyond these three examples might be
available.
\begin{itemize}
\item For $\nu_\mu$ events another signature is the life time of the
  resulting muon, see {\it e.g.}~\cite{LoSecco:1998bc,
    Brancus:2000we}: a $\mu^-$ can be captured by an atom to form a
  muonic atom and subsequently muon capture on the nucleus takes
  place. In this case, there will be no Michel electron. This process
  competes with ordinary muon decay, whereas for $\mu^+$ no such
  process is possible. The capture probability is
  approximately\footnote{There is a small correction to the lifetime
    of a captured $\mu^-$, due to the binding
    energy~\cite{Suzuki:1987jf}.}  given by the lifetime ratio
  $\tau_{\mu^-}/\tau_{\mu^+}$, where $\tau_{\mu^+}$ is the vacuum
  lifetime of $2.197\,\mu\mathrm{s}$~\cite{PDBook}. $\mu^-$ life times
  in common detector materials are~\cite{Suzuki:1987jf}:
  $2.026\,\mu\mathrm{s}$ for Carbon, {\it i.e.} liquid
  scintillator\footnote{The muon capture rate on Hydrogen is
    negligibly small.}, yielding a capture probability of 8\%;
  $1.795\,\mu\mathrm{s}$ for Oxygen, {\it i.e.}
  water\footnotemark[\thefootnote], yielding a capture probability of
  18\%; $0.537\,\mu\mathrm{s}$ for Argon, yielding a capture
  probability of 76\%. This effect has been used by the Kamiokande
  collaboration to determine the charge ratio of cosmic ray muons with
  an accuracy of 6\%~\cite{Yamada:1991aq}.  Here problems can arise
  due to the need, at least in some detectors like a WC, to positively
  identify the muon decay in order to distinguish the muon from a
  pion. For these detectors, the effect would be a reduced efficiency
  for $\nu$ events compared to $\bar\nu$ events. On the other hand,
  detectors which do not require the muon decay as particle
  identification tag, $\bar\nu_\mu$ charged current events which lead
  to muon capture, {\it i.e.} have no Michel electron, would constitute a
  very clean sample of $\bar\nu$ events. In the case of LAr, this
  sample would have an efficiency of about $0.5-0.6$.
\item Another difference between $\nu$ and $\bar\nu$ QE events is the
  distribution of $\cos\theta$, where $\theta$ is the angle between
  the incoming neutrino and the outgoing lepton in the laboratory
  frame.  Therefore, fitting the angular distribution of the charged
  leptons from QE events with respect to the neutrino beam direction
  provides a statistical handle on the $\nu/\bar\nu$ content of the
  beam. The MiniBooNE collaboration reports that they can use this
  effect in combination with the muon life time to determine a
  neutrino contamination of their anti-neutrino beam of 30\% with an
  accuracy of better than 10\%, {\it i.e.} the error in subtracting
  the neutrino background relative to all events is of the order
  3\%~\cite{Wascko:2006ty}. The difference in angular distribtuions is
  largest for neutrino energies around $1\,\mathrm{GeV}$ and is
  somewhat smaller at those energies we are looking at. Thus, this
  discriminant most likely has to be used in combination with other
  techniques.
\item Finally, the outgoing nucleon from a QE interaction is different
  for neutrino and anti-neutrino events: a proton for a $\nu$ event
  and a neutron for a $\bar\nu$ event, see equation~\ref{eq:QE}.
  Tagging the proton (being a charged particle) requires a
  sufficiently low energy threshold and sufficient spatial resolution
  to uniquely identify the proton track. Clearly, a liquid Argon
  detector fulfills both these conditions~\cite{Arneodo:2006ug}. On
  the other hand the proton tagging efficiency in water is very low,
  due to the Cerenkov threshold~\cite{Beacom:2003zu}. Tagging the
  neutron can be achieved by observing neutron capture onto a
  sufficiently heavy nucleus, which in turn will emit a
  $\gamma$-cascade with a total energy release of several MeV. The
  problem here is the competition between capture on light nuclei,
  which produces too little energy in $\gamma$-rays, and heavy nuclei.
  For a water Cerenkov detector the addition of a about 0.2\% of
  Gadolinium would allow to tag neutrons with an efficiency of about
  90\%~\cite{Beacom:2003nk}. Apart from the proton/neutron detection
  efficiency, charge exchange reactions where a proton becomes a
  neutron or {\it vice versa} would limit the achievable purity of
  this tag. Especially, since most detectors will be only able to tag
  either neutrons or protons and not both. The K2K collaboration has
  reported~\cite{Walter:2002sa} that about 70\% of nucleons in a
  quasi-elastic charged current events leave the nucleus without
  further interaction. The energy range of incoming neutrinos is
  $0.5-3.5\,\mathrm{GeV}$, {\it i.e.} close to the energies considered
  here. The remaining 30\% of events have the nucleon undergo elastic
  scattering inside the nucleus. Production of pions due to
  re-interactions happens only for proton momenta in excess of
  $1\,\mathrm{GeV}$, which is a small fraction of the overall events.
  Assuming an iso-scalar target, the probability to hit a neutron is
  0.5; further, assuming that in all elastic collisions full energy
  transfer between projectile and target takes place, we obtain that
  $0.5\cdot0.3=0.15$ of all events undergo a charge exchange. Thus
  purities at the level of 80\% seem possible using this technique.
\end{itemize}

These examples indicate that at least a statistical separation of
$\nu$ and $\bar\nu$ events seems possible without the use of magnetic
fields.  While we do not claim that any of these methods has been
proved to work with sufficient accuracy for our purposes, the
obtainable efficiencies and purities seem reasonably high to merit a
detailed investigation. In the following we will consider the impact
of various levels of statistical $\nu/\bar\nu$ separation on the
obtainable physics sensitivities, with the hope that our results will
trigger dedicated studies on statistical $\nu/\bar\nu$ separation in
different detectors. Therefore, we will resort to a highly idealized
parametrization of statistical separation of $\nu$ and $\bar\nu$,
which nevertheless is sufficient to illustrate the principle. We group
all events into two samples $N_1$ and $N_2$, which will be a mixture
of neutrino $N_\nu$ and anti-neutrino events $N_{\bar\nu}$:
\begin{eqnarray}
\label{eq:param}
N_1^i&=&\frac{1-p}{2} N_\nu^i + \frac{1+p}{2} N_{\bar\nu}^i\,\nonumber\\
N_2^i&=&\frac{1+p}{2} N_\nu^i + \frac{1-p}{2} N_{\bar\nu}^i\,,
\end{eqnarray}
where $p$ is the separation coefficient ($0\le p \le 1$), and $i$
labels the energy bins. A value $p=0$ is equivalent to no separation
at all, whereas $p=1$ stands for perfect separation.  Thus for $p\sim
1$, $N_1$ contains more anti-neutrino events and $N_2$ more neutrino
events. In some sense, $(1+p)/2$ is the efficiency of the separation
and $(1-p)/2$ is the contamination of the sample.  Clearly, in a real
detector efficiency and contamination need not add up to 1, nor need
the anti-neutrino efficiency and contamination in sample $N_1$ be the
same as the neutrino efficiency and contamination in sample $N_2$.
Furthermore, in general one expects that $p$ depends on the neutrino
energy (and hence on the index $i$ in equation~\ref{eq:param}), an
effect we neglect here. Furthermore, we assume that $p$ has been
determined by the near detector complex of the neutrino factory with
negligible errors.

Note, that in principle, polarization of the initial muons can serve a
similar purpose, {\it i.e.}\ improving the ratio of wrong sign to
right sign muons. From initial estimates it seems that a muon
polarization of about 50\% is equivalent to a value of
$p\simeq0.2-0.3$. Thus it may not be sufficient on its own, since 50\%
polarization is already quite ambitious~\cite{Blondel:2000vz}, but in
combination with the other techniques mentioned above it could be very
useful.

In this letter we will neglect all possible backgrounds, like neutral
current or charged current events with a leading pion. This
approximation can be justified by looking at the statistical error
derived from equation~\ref{eq:param}. For the signal being neutrinos
$N_\nu^i$ we obtain
\begin{equation}
\sigma_\mathrm{stat}^2=\frac{1+p}{2} N_\nu^i + \frac{1-p}{2}
N_{\bar\nu}^i\ + \frac{B_i}{2}\qquad
\stackrel{N_\nu^i\rightarrow0}{\longrightarrow}\qquad 
\frac{1-p}{2}N_{\bar\nu}^i\ + \frac{B_i}{2}\,,
\end{equation}
where $B_i$ is the background in bin $i$. The factor $1/2$ for $B_i$
arises from the assumption that the background is equally divided
between the samples $N_1$ and $N_2$, {\it i.e.} no $\nu/\bar\nu$
separation is applied. Thus, for $B_i\lesssim N_{\bar\nu}^i$ the
effect of the background will be small.  To conservatively estimate
the permissible background fraction we will assume that all
backgrounds migrate from the bin containing the most right sign
neutrinos $N_\mathrm{max}$ into that bin which contains the least
right sign neutrinos $N_\mathrm{min}$. The ratio
$r=N_\mathrm{min}/N_\mathrm{max}$ is $r\sim1/100$ for the energy
resolution of a TASD or LAr detector, {\it c.f.}\
figure~\ref{fig:spec}(a), and it is $r\sim1/10$ for the energy
resolution of a WC. The maximally allowable background fraction is
thus given by $r(1-p)$, which translates into a range of $0.001-0.003$
for TASD and LAr and $0.03-0.1$ for WC. These levels of background
rejection are within the margins of the current understanding of these
detectors, see {\it e.g}
\cite{Geer:2007kn,Yanagisawa:2007zz,Barger:2007yw}. In any case, a
full detector simulation with a special emphasize on nuclear effects
will be required to obtain a quantitatively reliable result for both
the obtainable background fraction and $\nu/\bar\nu$ separation.

\section{Sensitivity calculations}
\label{sec:sensitivities}

\begin{table}
\begin{tabular}{l@{\quad}|@{\quad}c@{\quad}c@{\quad}c}
  \hline \hline 
  & TASD~\cite{Geer:2007kn,Bross:2007ts} 
  & WC~\cite{Barger:2006vy} 
  & LAr~\cite{Barger:2007yw}\\ 
\hline 
  fiducial mass [kt] & 20 & 500 & 100\\ 
  efficiency & 0.73 & 0.9\footnote{on top of the single ring
  selection efficiency and an efficiency of 82\% for
  $\nu_\mu$ events} & 0.8\\ 
  magnetized &yes & no & no\\ 
  $\Delta E$ at $2.5\,\mathrm{GeV}$ [MeV] & 165 &
  300\footnote{equivalent Gau\ss ian width} & 165\\ 
  $p$ for muons & 0.999 & $0-0.7$ & $0.7-0.9$ \\ 
  $p$ for electrons & 0 & 0 & $0.7-0.9$\\ 
\hline \hline
\end{tabular}
\mycaption{\label{tab:det} Summary of relevant detector parameters.
Further details of our simulations can be found the references given
in the first line of the table.}
\end{table}

For the following results we considered three types of detectors: a
totally active magnetized scintillator detector
(TASD)~\cite{Geer:2007kn}, a megaton scale water Cerenkov (WC)
detector~\cite{Nakamura:2003hk,Jung:1999jq,deBellefon:2006vq}, and a
liquid Argon time projection chamber
(LAr)~\cite{Ereditato:2005ru}. Our TASD has similar properties to the
detector considered in~\cite{Geer:2007kn,Bross:2007ts} and it will
serve as benchmark setup for the performance of a low energy neutrino
factory. For the purposes of this letter, the main difference between
different detector technologies is mainly given by the energy
resolution for QE events, the attainable fiducial mass and whether
they can be magnetized. The relevant detector properties are
summarized in table~\ref{tab:det}; the simulations follow the details
given in the references shown in the table.

For both, the TASD and LAr we assume that QE and non-QE events can be
separated and we parametrize the energy resolution as $\Delta E =
r\sqrt{E} + 0.085$ in units of GeV, with $r = 0.05$ for QE events for
both, TASD and LAr, and $r = 0.2\,(0.3)$ for non-QE events for LAr
(TASD). For the TASD we assume charge identification at the level of
$10^{-3}$ for muons~\cite{Geer:2007kn}, and hence we take $p =
0.999$. We do include also $e$-like events in the TASD without charge
identification. In the case of LAr we assume that $\nu/\bar\nu$
separations in the range $0.7 \lesssim p \lesssim 0.9$ can be obtained
for $\mu$-like and $e$-like QE events; non-QE events are included
without $\nu/\bar\nu$ separation ($p=0$).
For the WC we use only single ring events, and the energy resolution
is obtained from a full simulation based on the SuperK Monte Carlo
taken from~\cite{Yanagisawa:2007zz}, including the contribution of
non-QE events which pass the single ring criterion. We account for the
fact that for captured $\mu^-$ no Michel electron can be observed by
an additional efficiency of 82\% for $\nu_\mu$ events. We consider
$\nu/\bar\nu$ separations in the range $0 \le p \lesssim 0.7$ for
$\mu$-like events. Although some of the separation methods mentioned
above might work also for $\nu_e$ events ($\cos\theta$ distribution
and neutron tagging), we conservatively assume here no $\nu/\bar\nu$
separation for $e$-like events in a WC.

For the neutrino factory we use a stored muon energy $E_\mu$ of
$5\,\mathrm{GeV}$~\footnote{We have verified that this energy is close
to optimal for the baseline considered here, in agreement
with~\cite{Huber:2007uj}.}  and total of $10^{22}$ useful muon decays,
equally divided into $\mu^-$ and $\mu^+$ running. This luminosity
corresponds to 10 years total running time of the baseline setup of
the International Design Study for a neutrino
factory~\cite{idsbase}.\footnote{This setup assumes a $4\,\mathrm{MW}$
proton beam, for $10^7\,\mathrm{s}$ a year.  Fermilab's project X will
deliver $2.3\,\mathrm{MW}$ of protons for $1.7\cdot10^7\,\mathrm{s}$
per year. As a result the expect neutrino luminosity per calendar year
should be approximately the same.}  We assume a baseline of
$1\,290\,\mathrm{km}$, which corresponds to the distance from Fermilab
to the Deep Underground Science and Engineering Laboratory (DUSEL) at
Homestake.
For the sake of comparison with conventional neutrino beams we also
will show results for a $500\,\mathrm{kt}$ WC in a wide-band neutrino
beam stemming from $120\,\mathrm{GeV}$ protons with the same baseline
($1\,290\,\mathrm{km}$) and at an off-axis angle of
$58\,\mathrm{mrad}$. The beam power (4~MW) and the running time
(10~yr) is assumed to be the same as for the neutrino factory. This
corresponds (except for the larger detector mass) to the setup
considered in~\cite{Barger:2006vy} and will be labeled as WBB.

To calculate the sensitivities we will use $\Delta
m^2_{31}=2.5\cdot10^{-3}\,\mathrm{eV}^2$, $\sin^2\theta_{23}=0.5$,
$\Delta m^2_{21}=7.6\cdot10^{-5}\,\mathrm{eV}^2$ and
$\sin^2\theta_{12}=0.3$, which corresponds to the results found in
version 6 of~\cite{Maltoni:2004ei}. For $\theta_{13}$ and $\delta$ we
assume that they have to be determined by the experimental setups
considered. The analysis is performed with
GLoBES~\cite{Huber:2004ka,Huber:2007ji} using a 4\% error on the solar
parameters $\Delta m^2_{21}$ and $\sin^2\theta_{12}$ and a 5\% error
on the matter density. We impose no external information on $\Delta
m^2_{31}$ and $\theta_{23}$ since these parameters are measured by the
considered experiment with good precision. We always assume a true
normal neutrino mass hierarchy, but we have checked that your results
are not significantly changed when the mass hierarchy is inverted. We
assume a 2.5\% systematic error on each signal. All sensitivities are
evaluated at the $3\,\sigma$ confidence level for 1 degree of freedom,
{\it i.e.}  $\Delta\chi^2=9$.

\begin{figure}
\centering \includegraphics[width=0.55\textwidth]{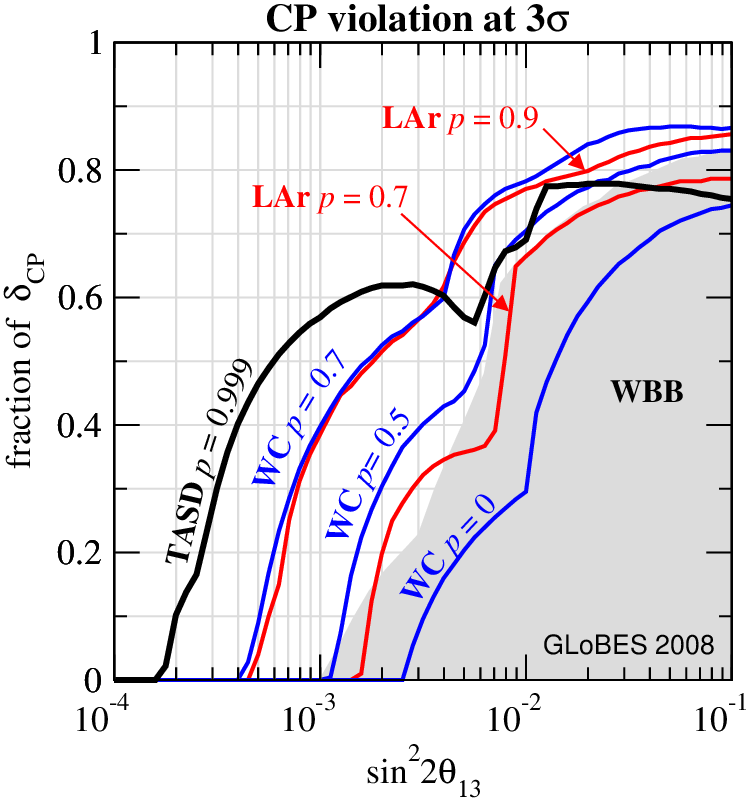}
  \mycaption{\label{fig:cpf} Fraction of $\delta$ for which CP
  violation can be discovered at $3\,\sigma$ confidence level for
  different experiments as described in table~\ref{tab:det}. The
  numbers next to the lines correspond to different values of the
  $\nu/\bar\nu$ separation coefficient $p$ as defined in
  equation~\ref{eq:param}. The shaded region corresponds to the WBB.}
\end{figure}

In figure~\ref{fig:cpf} we show the obtainable sensitivities to CP
violation as a function of the true value of $\sin^22\theta_{13}$ for
the different detectors as described in table~\ref{tab:det}. First, we
note that the conventional WBB setup performs very well for large
values of $\sin^22\theta_{13}>0.03$. For
$0.006<\sin^22\theta_{13}<0.03$, a low energy neutrino factory with a
magnetized TASD performs marginally better than a WBB and only for
$\sin^22\theta_{13}<0.006$ the neutrino factory yields a considerable
improvement in sensitivity. A WC with $p=0$, {\it i.e.} no
$\nu/\bar\nu$ separation at all, will perform worse in a neutrino
factory beam than in a wide band beam. However, already for a modest
separation of $p=0.5$, the WC would have the same or even better
performance than a TASD for $\sin^22\theta_{13}>0.006$. For good
$\nu/\bar\nu$ separation, $p=0.7$, the WC outperforms a TASD down to
$\sin^22\theta_{13}>0.004$. For LAr the better energy resolution
largely allows to compensate the smaller mass and for a somewhat
larger value of $p=0.9$ it is more or less equivalent to the WC with
$p=0.7$. These results clearly demonstrate that non-magnetized
detectors can exploit their relatively larger mass compared to
magnetized ones in order to address the same physics in a low energy
neutrino factory beam. The question which technology yields better
sensitivities depends on the value of $\sin^22\theta_{13}$, the degree
of $\nu/\bar\nu$ separation and the relative detector mass.

\begin{figure}
\centering
  \includegraphics[width=0.9\textwidth]{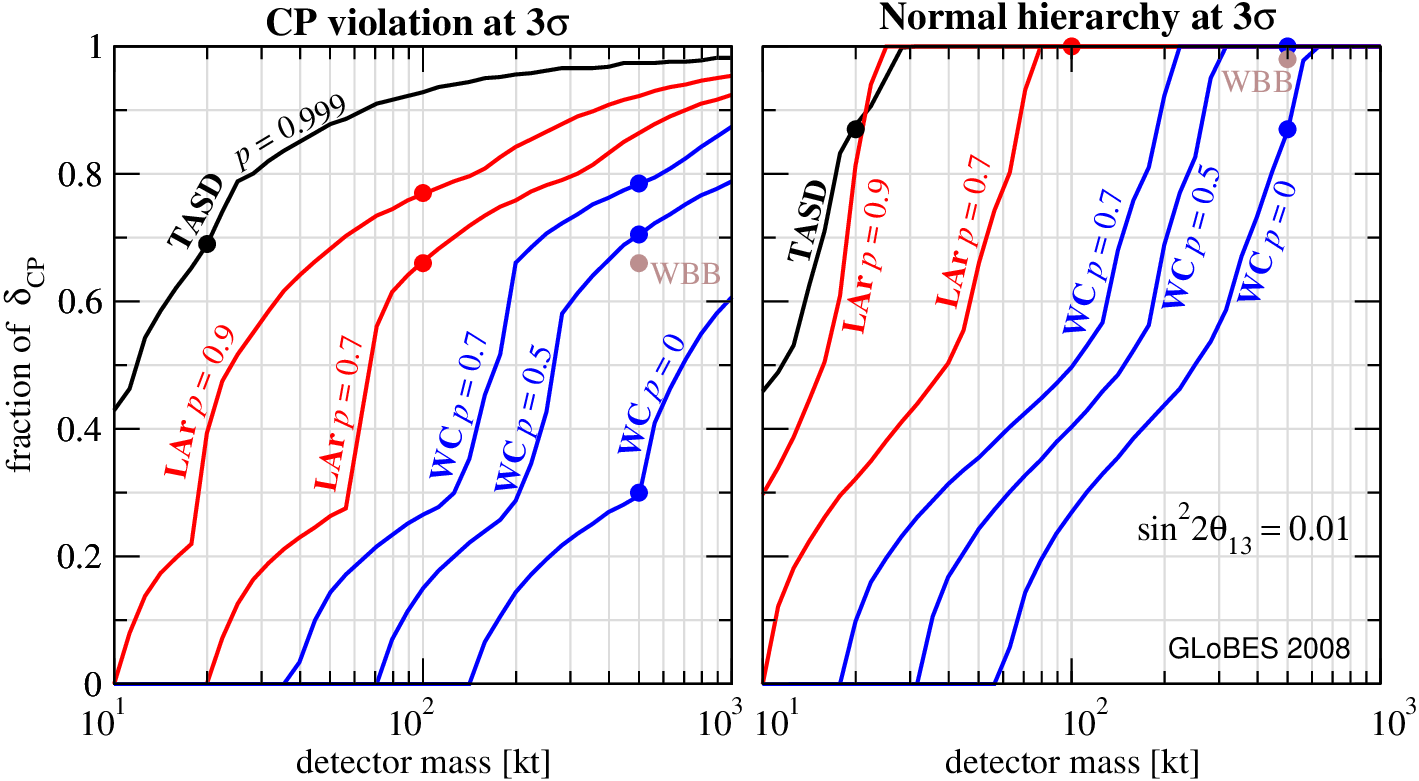}
  \mycaption{\label{fig:lumi} Fraction of $\delta$ as function of the
  detector mass for which CP violation (left hand panel) or the mass
  hierarchy can be discovered (right hand panel) at $3\,\sigma$
  confidence level for different experiments as described in
  table~\ref{tab:det} for $\sin^22\theta_{13} = 0.01$. The numbers
  next to the lines correspond to different values of the
  $\nu/\bar\nu$ separation coefficient $p$ as defined in
  equation~\ref{eq:param}. }
\end{figure}

Therefore, we study the physics reach as a function of the detector
mass. This is shown in figure~\ref{fig:lumi} for a true value of
$\sin^22\theta_{13}=0.01$. The left hand panel shows the fraction of
$\delta$ for which CP violation can be discovered, whereas the right
hand panel shows the fraction of $\delta$ for which a normal mass
hierarchy can be identified. The dots indicate the sensitivity
obtained for the detector masses as specified in
table~\ref{tab:det}. From the right hand panel it is obvious that the
determination of the mass hierarchy can be achieved by any technology
for almost all values of the CP phase. Let us note that for the
hierarchy determination $\nu_e$ events contribute significantly to the
sensitivity, even with $p=0$, and this contribution is further enhanced
if some $\nu/\bar\nu$ separation is assumed also for $e$-like events
(see~\cite{Schwetz:2007py} for an explanation). This is important also
for the CP violation measurement, since the hierarchy degenerate
solution often is located at CP conserving values of $\delta$. Indeed,
the kink visible in the curves shown in the left panel, above which
the sensitivity improves drastically, corresponds roughly to the
detector mass for which the sign degeneracy can be lifted. Therefore,
the inclusion of electron events (and increasing $p$ for them) shifts
this kink to lower detector masses; though it has very little impact
on the CP sensitivity at high luminosities, which is dominated by
$\mu$-like events.

The left hand panel shows that, depending on the
detector type and level of $\nu/\bar\nu$ separation, a larger detector
mass is needed to achieve the same sensitivity as the usual magnetized
TASD with a fiducial mass of $20\,\mathrm{kt}$. For the WC, we find
equivalent masses in the range from $200-500\,\mathrm{kt}$ for
$p=0.7-0.5$ and for the LAr the mass range is from
$50-110\,\mathrm{kt}$ for $p=0.9-0.7$. The equivalent masses increase
for smaller values of $\theta_{13}$ and for
$\sin^22\theta_{13}=0.003$, the equivalent mass ranges become
$m=500-900\,\mathrm{kt}$ for WC and $m=110-300\,\mathrm{kt}$ for LAr.

So far we have assumed maximal mixing $\theta_{23}=\pi/4$. Let us now
investigate the impact of non-maximal values for $\theta_{23}$ on our
results. Similar to a finite energy resolution also non-maximal values
of $\theta_{23}$ will lead to a wrong sign muon background at the
first oscillation maximum, since the survival probability $P_{\mu\mu}$
goes not to zero.  In the example shown in figure~\ref{fig:spec}(a),
the background from the energy resolution is about 10 events per
bin. The unoscillated event rate in that bin would be about 300
events, thus we have a background suppression by about a factor of 30
for $\theta_{23} = \pi/4$. We can estimate the excursion of
$\theta_{23}$ from maximality which would cause the same level of
events by solving $1-\sin^22\theta_{23}=1/30$.  We find that
$\theta_{23}\simeq\pi/4\pm0.1$ satisfies this constraint; this is
equivalent to a variation of $\sin^2\theta_{23}=0.5\pm0.1$, which is
about the $2\,\sigma$ range currently allowed by global neutrino
data~\cite{Maltoni:2004ei}. Since the significance of the signal is
due to not only one bin at $E_\mathrm{1st}$, but due to the cumulative
effect of many bins close by, which experience reduction of right sign
muon events much smaller than 30, one can expect that the proposed
scheme will not be spoiled by reasonable deviations from maximal
mixing.

\begin{figure}
\centering 
  \includegraphics[width=0.45\textwidth]{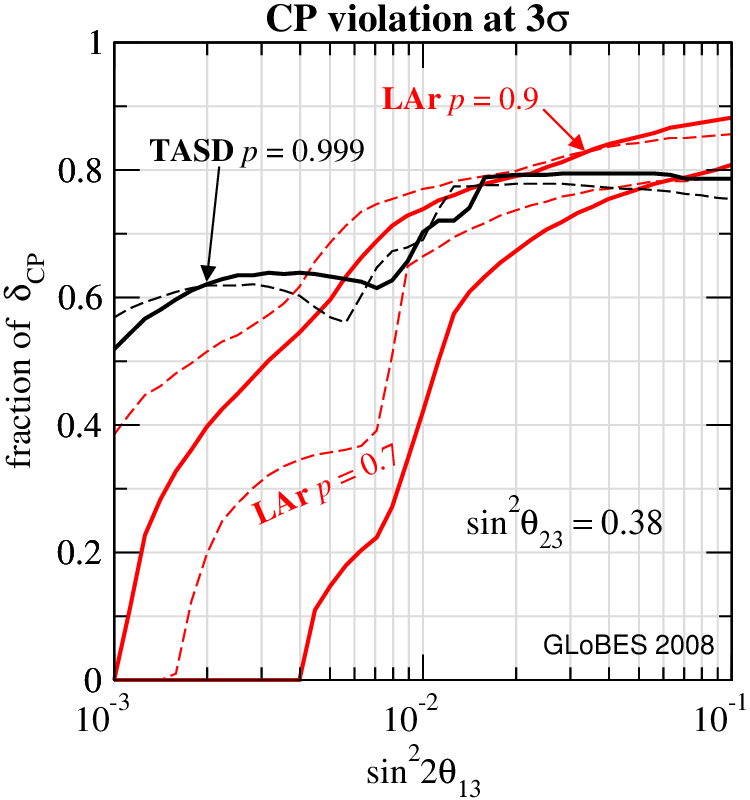}
  \includegraphics[width=0.45\textwidth]{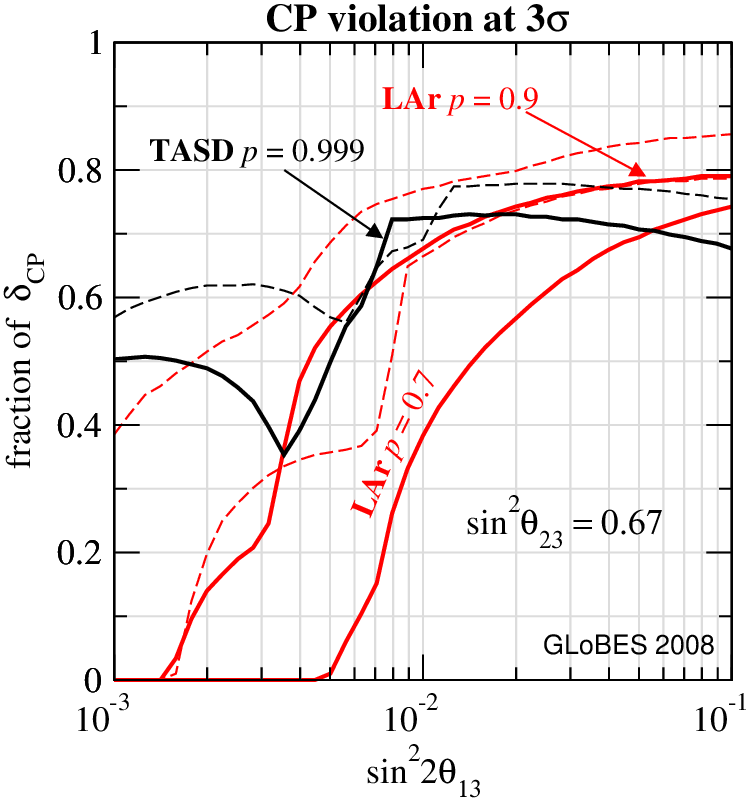}
  \mycaption{\label{fig:non-max} Fraction of $\delta$ for which CP
  violation can be discovered at $3\,\sigma$ confidence level as a
  function of $\sin^22\theta_{13}$ for the TASD and LAr ($p = 0.7$ and
  $0.9$) setups from table~\ref{tab:det}, for
  $\sin^2\theta_{23} = 0.38$ (left) and $\sin^2\theta_{23} = 0.67$
  (right). For comparison we show also the CP fractions
  for $\sin^2\theta_{23} = 0.5$ (dashed curves).}
\end{figure}

Figure~\ref{fig:non-max} shows the sensitivity to CP violation for the
LAr detector compared to the magnetized TASD for the
current~\cite{Maltoni:2004ei} lower $2\sigma$ bound (left panel) and
upper $3\sigma$ bound (right panel) on $\sin^2\theta_{23}$. As
expected we find a somewhat worse sensitivity for non-maximal values,
however the relative performance of the magnetic and non-magnetic
detectors is similar to maximal mixing. Note that the CP signal itself
becomes smaller for $\theta_{23} \neq \pi/4$ since it is proportional
to $\sin2\theta_{23}$, {\it c.f.}\ equation~\ref{eq:app}. We conclude
that for reasonably non-maximal values of $\theta_{23}$ our results
are not significantly affected.

\section{Conclusions}
\label{sec:conclusions}

The results presented in this letter show that a sufficiently well
performing non-magnetized detector may be able to cover the physics
needs of a low energy neutrino factory for $\sin^22\theta_{13}$ larger
than about $10^{-3}$. Detector requirements are a statistical
neutrino/anti-neutrino separation at the level of 50\% to 90\%, a good
energy resolution, and large fiducial masses in the range of 100 to
500~kt. In this way, a neutrino factory beam does not {\it a priori}
exclude the use of multi-purpose detectors, which have other
interesting applications in astrophysics or proton decay.
Furthermore, a low energy neutrino factory exploiting an already
existing, large non-magnetized detector can serve as intermediate step
between the super beam program and a full scale, high energy neutrino
factory.  We hope that the results presented here will stimulate a
detailed investigation of the required detector capabilities.

\acknowledgments

We would like to thank John Beacom, Takaaki Kajita, Jonathan Link,
Mauro Mezzetto, Andr{\'e} Rubbia and Mark Vagins for useful
discussions. We acknowledge the support of the European
Community-Research Infrastructure Activity under the FP6 ``Structuring
the European Research Area'' program (CARE, contract number
RII3-CT-2003-506395).

\bibliographystyle{apsrev}
\bibliography{./paper}

\begin{thebibliography}{35}
\expandafter\ifx\csname natexlab\endcsname\relax\def\natexlab#1{#1}\fi
\expandafter\ifx\csname bibnamefont\endcsname\relax
  \def\bibnamefont#1{#1}\fi
\expandafter\ifx\csname bibfnamefont\endcsname\relax
  \def\bibfnamefont#1{#1}\fi
\expandafter\ifx\csname citenamefont\endcsname\relax
  \def\citenamefont#1{#1}\fi
\expandafter\ifx\csname url\endcsname\relax
  \def\url#1{\texttt{#1}}\fi
\expandafter\ifx\csname urlprefix\endcsname\relax\def\urlprefix{URL }\fi
\providecommand{\bibinfo}[2]{#2}
\providecommand{\eprint}[2][]{\url{#2}}

\bibitem[{\citenamefont{Geer}(1998)}]{Geer:1997iz}
\bibinfo{author}{\bibfnamefont{S.}~\bibnamefont{Geer}}, \bibinfo{journal}{Phys.
  Rev.} \textbf{\bibinfo{volume}{D57}}, \bibinfo{pages}{6989}
  (\bibinfo{year}{1998}), \eprint{hep-ph/9712290}.

\bibitem[{\citenamefont{Berg et~al.}(2008{\natexlab{a}})}]{issacc:2008xx}
\bibinfo{author}{\bibfnamefont{J.~S.} \bibnamefont{Berg}} \bibnamefont{et~al.}
  (\bibinfo{collaboration}{ISS Accelerator Working Group})
  (\bibinfo{year}{2008}{\natexlab{a}}), \eprint{0802.4023}.

\bibitem[{\citenamefont{Abe et~al.}(2007)}]{Abe:2007bi}
\bibinfo{author}{\bibfnamefont{T.}~\bibnamefont{Abe}} \bibnamefont{et~al.}
  (\bibinfo{collaboration}{ISS Detector Working Group}) (\bibinfo{year}{2007}),
  \eprint{0712.4129}.

\bibitem[{\citenamefont{Bandyopadhyay et~al.}(2007)}]{Bandyopadhyay:2007kx}
\bibinfo{author}{\bibfnamefont{A.}~\bibnamefont{Bandyopadhyay}}
  \bibnamefont{et~al.} (\bibinfo{collaboration}{ISS Physics Working Group})
  (\bibinfo{year}{2007}), \eprint{0710.4947}.

\bibitem[{\citenamefont{Geer et~al.}(2007)\citenamefont{Geer, Mena, and
  Pascoli}}]{Geer:2007kn}
\bibinfo{author}{\bibfnamefont{S.}~\bibnamefont{Geer}},
  \bibinfo{author}{\bibfnamefont{O.}~\bibnamefont{Mena}}, \bibnamefont{and}
  \bibinfo{author}{\bibfnamefont{S.}~\bibnamefont{Pascoli}},
  \bibinfo{journal}{Phys. Rev.} \textbf{\bibinfo{volume}{D75}},
  \bibinfo{pages}{093001} (\bibinfo{year}{2007}), \eprint{hep-ph/0701258}.

\bibitem[{\citenamefont{Bross et~al.}(2007)\citenamefont{Bross, Ellis, Geer,
  Mena, and Pascoli}}]{Bross:2007ts}
\bibinfo{author}{\bibfnamefont{A.~D.} \bibnamefont{Bross}},
  \bibinfo{author}{\bibfnamefont{M.}~\bibnamefont{Ellis}},
  \bibinfo{author}{\bibfnamefont{S.}~\bibnamefont{Geer}},
  \bibinfo{author}{\bibfnamefont{O.}~\bibnamefont{Mena}}, \bibnamefont{and}
  \bibinfo{author}{\bibfnamefont{S.}~\bibnamefont{Pascoli}}
  (\bibinfo{year}{2007}), \eprint{0709.3889}.

\bibitem[{\citenamefont{Huber and Winter}(2007)}]{Huber:2007uj}
\bibinfo{author}{\bibfnamefont{P.}~\bibnamefont{Huber}} \bibnamefont{and}
  \bibinfo{author}{\bibfnamefont{W.}~\bibnamefont{Winter}},
  \bibinfo{journal}{Phys. Lett.} \textbf{\bibinfo{volume}{B655}},
  \bibinfo{pages}{251} (\bibinfo{year}{2007}), \eprint{0706.2862}.

\bibitem[{\citenamefont{Drakoulakos et~al.}(2004)}]{Minerva}
\bibinfo{author}{\bibfnamefont{D.}~\bibnamefont{Drakoulakos}}
  \bibnamefont{et~al.} (\bibinfo{collaboration}{Minerva})
  (\bibinfo{year}{2004}), \eprint{hep-ex/0405002}.

\bibitem[{\citenamefont{Freund et~al.}(2000)\citenamefont{Freund, Huber, and
  Lindner}}]{Freund:2000ti}
\bibinfo{author}{\bibfnamefont{M.}~\bibnamefont{Freund}},
  \bibinfo{author}{\bibfnamefont{P.}~\bibnamefont{Huber}}, \bibnamefont{and}
  \bibinfo{author}{\bibfnamefont{M.}~\bibnamefont{Lindner}},
  \bibinfo{journal}{Nucl. Phys.} \textbf{\bibinfo{volume}{B585}},
  \bibinfo{pages}{105} (\bibinfo{year}{2000}), \eprint{hep-ph/0004085}.

\bibitem[{\citenamefont{Aoki et~al.}(2005)\citenamefont{Aoki, Hagiwara, and
  Okamura}}]{Aoki:2003kc}
\bibinfo{author}{\bibfnamefont{M.}~\bibnamefont{Aoki}},
  \bibinfo{author}{\bibfnamefont{K.}~\bibnamefont{Hagiwara}}, \bibnamefont{and}
  \bibinfo{author}{\bibfnamefont{N.}~\bibnamefont{Okamura}},
  \bibinfo{journal}{Phys. Lett.} \textbf{\bibinfo{volume}{B606}},
  \bibinfo{pages}{371} (\bibinfo{year}{2005}), \eprint{hep-ph/0311324}.

\bibitem[{\citenamefont{Autiero et~al.}(2007)}]{Autiero:2007zj}
\bibinfo{author}{\bibfnamefont{D.}~\bibnamefont{Autiero}} \bibnamefont{et~al.},
  \bibinfo{journal}{JCAP} \textbf{\bibinfo{volume}{0711}}, \bibinfo{pages}{011}
  (\bibinfo{year}{2007}), \eprint{0705.0116}.

\bibitem[{\citenamefont{{Yao} et~al.}(2006)}]{PDBook}
\bibinfo{author}{\bibfnamefont{W.-M.} \bibnamefont{{Yao}}}
  \bibnamefont{et~al.}, \bibinfo{journal}{{Journal of Physics G}}
  \textbf{\bibinfo{volume}{33}}, \bibinfo{pages}{1+} (\bibinfo{year}{2006}),
  \urlprefix\url{http://pdg.lbl.gov}.

\bibitem[{\citenamefont{Paschos and Yu}(2002)}]{Paschos:2001np}
\bibinfo{author}{\bibfnamefont{E.~A.} \bibnamefont{Paschos}} \bibnamefont{and}
  \bibinfo{author}{\bibfnamefont{J.~Y.} \bibnamefont{Yu}},
  \bibinfo{journal}{Phys. Rev.} \textbf{\bibinfo{volume}{D65}},
  \bibinfo{pages}{033002} (\bibinfo{year}{2002}), \eprint{hep-ph/0107261}.

\bibitem[{\citenamefont{Maltoni et~al.}(2004)\citenamefont{Maltoni, Schwetz,
  Tortola, and Valle}}]{Maltoni:2004ei}
\bibinfo{author}{\bibfnamefont{M.}~\bibnamefont{Maltoni}},
  \bibinfo{author}{\bibfnamefont{T.}~\bibnamefont{Schwetz}},
  \bibinfo{author}{\bibfnamefont{M.~A.} \bibnamefont{Tortola}},
  \bibnamefont{and} \bibinfo{author}{\bibfnamefont{J.~W.~F.}
  \bibnamefont{Valle}}, \bibinfo{journal}{New J. Phys.}
  \textbf{\bibinfo{volume}{6}}, \bibinfo{pages}{122} (\bibinfo{year}{2004}),
  \eprint{hep-ph/0405172}.

\bibitem[{\citenamefont{LoSecco}(1999)}]{LoSecco:1998bc}
\bibinfo{author}{\bibfnamefont{J.~M.} \bibnamefont{LoSecco}},
  \bibinfo{journal}{Phys. Rev.} \textbf{\bibinfo{volume}{D59}},
  \bibinfo{pages}{117302} (\bibinfo{year}{1999}), \eprint{hep-ph/9806318}.

\bibitem[{\citenamefont{Brancus et~al.}(2000)}]{Brancus:2000we}
\bibinfo{author}{\bibfnamefont{I.~M.} \bibnamefont{Brancus}}
  \bibnamefont{et~al.}, \bibinfo{journal}{Acta Phys. Polon.}
  \textbf{\bibinfo{volume}{B31}}, \bibinfo{pages}{465} (\bibinfo{year}{2000}).

\bibitem[{\citenamefont{Suzuki et~al.}(1987)\citenamefont{Suzuki, Measday, and
  Roalsvig}}]{Suzuki:1987jf}
\bibinfo{author}{\bibfnamefont{T.}~\bibnamefont{Suzuki}},
  \bibinfo{author}{\bibfnamefont{D.~F.} \bibnamefont{Measday}},
  \bibnamefont{and} \bibinfo{author}{\bibfnamefont{J.~P.}
  \bibnamefont{Roalsvig}}, \bibinfo{journal}{Phys. Rev.}
  \textbf{\bibinfo{volume}{C35}}, \bibinfo{pages}{2212} (\bibinfo{year}{1987}).

\bibitem[{\citenamefont{Yamada et~al.}(1991)}]{Yamada:1991aq}
\bibinfo{author}{\bibfnamefont{M.}~\bibnamefont{Yamada}} \bibnamefont{et~al.},
  \bibinfo{journal}{Phys. Rev.} \textbf{\bibinfo{volume}{D44}},
  \bibinfo{pages}{617} (\bibinfo{year}{1991}).

\bibitem[{\citenamefont{Wascko}(2006)}]{Wascko:2006ty}
\bibinfo{author}{\bibfnamefont{M.~O.} \bibnamefont{Wascko}}
  (\bibinfo{collaboration}{MiniBooNE}), \bibinfo{journal}{Nucl. Phys. Proc.
  Suppl.} \textbf{\bibinfo{volume}{159}}, \bibinfo{pages}{79}
  (\bibinfo{year}{2006}), \eprint{hep-ex/0602051}.

\bibitem[{\citenamefont{Arneodo et~al.}(2006)}]{Arneodo:2006ug}
\bibinfo{author}{\bibfnamefont{F.}~\bibnamefont{Arneodo}} \bibnamefont{et~al.}
  (\bibinfo{collaboration}{ICARUS-Milano}), \bibinfo{journal}{Phys. Rev.}
  \textbf{\bibinfo{volume}{D74}}, \bibinfo{pages}{112001}
  (\bibinfo{year}{2006}), \eprint{physics/0609205}.

\bibitem[{\citenamefont{Beacom and Palomares-Ruiz}(2003)}]{Beacom:2003zu}
\bibinfo{author}{\bibfnamefont{J.~F.} \bibnamefont{Beacom}} \bibnamefont{and}
  \bibinfo{author}{\bibfnamefont{S.}~\bibnamefont{Palomares-Ruiz}},
  \bibinfo{journal}{Phys. Rev.} \textbf{\bibinfo{volume}{D67}},
  \bibinfo{pages}{093001} (\bibinfo{year}{2003}), \eprint{hep-ph/0301060}.

\bibitem[{\citenamefont{Beacom and Vagins}(2004)}]{Beacom:2003nk}
\bibinfo{author}{\bibfnamefont{J.~F.} \bibnamefont{Beacom}} \bibnamefont{and}
  \bibinfo{author}{\bibfnamefont{M.~R.} \bibnamefont{Vagins}},
  \bibinfo{journal}{Phys. Rev. Lett.} \textbf{\bibinfo{volume}{93}},
  \bibinfo{pages}{171101} (\bibinfo{year}{2004}), \eprint{hep-ph/0309300}.

\bibitem[{\citenamefont{Walter}(2002)}]{Walter:2002sa}
\bibinfo{author}{\bibfnamefont{C.~W.} \bibnamefont{Walter}},
  \bibinfo{journal}{Nucl. Phys. Proc. Suppl.} \textbf{\bibinfo{volume}{112}},
  \bibinfo{pages}{140} (\bibinfo{year}{2002}).

\bibitem[{\citenamefont{Blondel}(2000)}]{Blondel:2000vz}
\bibinfo{author}{\bibfnamefont{A.}~\bibnamefont{Blondel}},
  \bibinfo{journal}{Nucl. Instrum. Meth.} \textbf{\bibinfo{volume}{A451}},
  \bibinfo{pages}{131} (\bibinfo{year}{2000}).

\bibitem[{\citenamefont{Yanagisawa et~al.}(2007)\citenamefont{Yanagisawa, Jung,
  Le, and Viren}}]{Yanagisawa:2007zz}
\bibinfo{author}{\bibfnamefont{C.}~\bibnamefont{Yanagisawa}},
  \bibinfo{author}{\bibfnamefont{C.~K.} \bibnamefont{Jung}},
  \bibinfo{author}{\bibfnamefont{P.~T.} \bibnamefont{Le}}, \bibnamefont{and}
  \bibinfo{author}{\bibfnamefont{B.}~\bibnamefont{Viren}},
  \bibinfo{journal}{AIP Conf. Proc.} \textbf{\bibinfo{volume}{944}},
  \bibinfo{pages}{92} (\bibinfo{year}{2007}).

\bibitem[{\citenamefont{Barger et~al.}(2007)}]{Barger:2007yw}
\bibinfo{author}{\bibfnamefont{V.}~\bibnamefont{Barger}} \bibnamefont{et~al.}
  (\bibinfo{year}{2007}), \eprint{0705.4396}.

\bibitem[{\citenamefont{Barger et~al.}(2006)\citenamefont{Barger, Dierckxsens,
  Diwan, Huber, Lewis, Marfatia, and Viren}}]{Barger:2006vy}
\bibinfo{author}{\bibfnamefont{V.}~\bibnamefont{Barger}},
  \bibinfo{author}{\bibfnamefont{M.}~\bibnamefont{Dierckxsens}},
  \bibinfo{author}{\bibfnamefont{M.}~\bibnamefont{Diwan}},
  \bibinfo{author}{\bibfnamefont{P.}~\bibnamefont{Huber}},
  \bibinfo{author}{\bibfnamefont{C.}~\bibnamefont{Lewis}},
  \bibinfo{author}{\bibfnamefont{D.}~\bibnamefont{Marfatia}}, \bibnamefont{and}
  \bibinfo{author}{\bibfnamefont{B.}~\bibnamefont{Viren}},
  \bibinfo{journal}{Phys. Rev.} \textbf{\bibinfo{volume}{D74}},
  \bibinfo{pages}{073004} (\bibinfo{year}{2006}), \eprint{hep-ph/0607177}.

\bibitem[{\citenamefont{Nakamura}(2003)}]{Nakamura:2003hk}
\bibinfo{author}{\bibfnamefont{K.}~\bibnamefont{Nakamura}},
  \bibinfo{journal}{Int. J. Mod. Phys.} \textbf{\bibinfo{volume}{A18}},
  \bibinfo{pages}{4053} (\bibinfo{year}{2003}).

\bibitem[{\citenamefont{Jung}(2000)}]{Jung:1999jq}
\bibinfo{author}{\bibfnamefont{C.~K.} \bibnamefont{Jung}},
  \bibinfo{journal}{AIP Conf. Proc.} \textbf{\bibinfo{volume}{533}},
  \bibinfo{pages}{29} (\bibinfo{year}{2000}), \eprint{hep-ex/0005046}.

\bibitem[{\citenamefont{de~Bellefon et~al.}(2006)}]{deBellefon:2006vq}
\bibinfo{author}{\bibfnamefont{A.}~\bibnamefont{de~Bellefon}}
  \bibnamefont{et~al.} (\bibinfo{year}{2006}), \eprint{hep-ex/0607026}.

\bibitem[{\citenamefont{Ereditato and Rubbia}(2006)}]{Ereditato:2005ru}
\bibinfo{author}{\bibfnamefont{A.}~\bibnamefont{Ereditato}} \bibnamefont{and}
  \bibinfo{author}{\bibfnamefont{A.}~\bibnamefont{Rubbia}},
  \bibinfo{journal}{Nucl. Phys. Proc. Suppl.} \textbf{\bibinfo{volume}{154}},
  \bibinfo{pages}{163} (\bibinfo{year}{2006}), \eprint{hep-ph/0509022}.

\bibitem[{\citenamefont{Berg et~al.}(2008{\natexlab{b}})}]{idsbase}
\bibinfo{author}{\bibfnamefont{S.}~\bibnamefont{Berg}} \bibnamefont{et~al.},
  \bibinfo{type}{Tech. Rep.} \bibinfo{number}{IDS-NF-002},
  \bibinfo{institution}{IDS-NF} (\bibinfo{year}{2008}{\natexlab{b}}).

\bibitem[{\citenamefont{Huber et~al.}(2005)\citenamefont{Huber, Lindner, and
  Winter}}]{Huber:2004ka}
\bibinfo{author}{\bibfnamefont{P.}~\bibnamefont{Huber}},
  \bibinfo{author}{\bibfnamefont{M.}~\bibnamefont{Lindner}}, \bibnamefont{and}
  \bibinfo{author}{\bibfnamefont{W.}~\bibnamefont{Winter}},
  \bibinfo{journal}{Comput. Phys. Commun.} \textbf{\bibinfo{volume}{167}},
  \bibinfo{pages}{195} (\bibinfo{year}{2005}), \eprint{hep-ph/0407333}.

\bibitem[{\citenamefont{Huber et~al.}(2007)\citenamefont{Huber, Kopp, Lindner,
  Rolinec, and Winter}}]{Huber:2007ji}
\bibinfo{author}{\bibfnamefont{P.}~\bibnamefont{Huber}},
  \bibinfo{author}{\bibfnamefont{J.}~\bibnamefont{Kopp}},
  \bibinfo{author}{\bibfnamefont{M.}~\bibnamefont{Lindner}},
  \bibinfo{author}{\bibfnamefont{M.}~\bibnamefont{Rolinec}}, \bibnamefont{and}
  \bibinfo{author}{\bibfnamefont{W.}~\bibnamefont{Winter}},
  \bibinfo{journal}{Comput. Phys. Commun.} \textbf{\bibinfo{volume}{177}},
  \bibinfo{pages}{432} (\bibinfo{year}{2007}), \eprint{hep-ph/0701187}.

\bibitem[{\citenamefont{Schwetz}(2007)}]{Schwetz:2007py}
\bibinfo{author}{\bibfnamefont{T.}~\bibnamefont{Schwetz}},
  \bibinfo{journal}{JHEP} \textbf{\bibinfo{volume}{05}}, \bibinfo{pages}{093}
  (\bibinfo{year}{2007}), \eprint{hep-ph/0703279}.

\end{thebibliography}

\end{document}